\documentclass [superscriptaddress]{revtex4}
\usepackage{graphicx}
\usepackage{amssymb}
\usepackage{textcomp}

\begin{document}

\title{ Time-bin entangled photons from a quantum dot}

\author{Harishankar Jayakumar}
\email{harishankar@uibk.ac.at}
\affiliation{Institut f\"ur Experimentalphysik, Universit\"at Innsbruck, Technikerstrasse 25, 6020 Innsbruck, Austria}
\author{Ana Predojevi\'{c}}
\email{ana.predojevic@uibk.ac.at}
\affiliation{Institut f\"ur Experimentalphysik, Universit\"at Innsbruck, Technikerstrasse 25, 6020 Innsbruck, Austria}
\author{Thomas Kauten}
\affiliation{Institut f\"ur Experimentalphysik, Universit\"at Innsbruck, Technikerstrasse 25, 6020 Innsbruck, Austria}
\author{Tobias Huber}
\affiliation{Institut f\"ur Experimentalphysik, Universit\"at Innsbruck, Technikerstrasse 25, 6020 Innsbruck, Austria}
\author{Glenn S. Solomon}
\affiliation{Joint Quantum Institute, National Institute of Standards and Technology  \& University of Maryland, Gaithersburg, MD 20849, USA}
\author{Gregor Weihs}
\affiliation{Institut f\"ur Experimentalphysik, Universit\"at Innsbruck, Technikerstrasse 25, 6020 Innsbruck, Austria}
\maketitle

{\bfseries Long distance quantum communication \cite{Gisin} is one of the prime goals in the field of quantum information science \cite{Nielsen}. With information encoded in the quantum state of photons, existing telecommunication fiber networks can be effectively used as a transport medium. To achieve this goal, a source of robust entangled single photon pairs is required. While time-bin entanglement \cite{Brendel} offers the required robustness \cite{Honjo, Dynes}, currently used parametric down-conversion sources have limited performance due to multi-pair contributions. We report the realization of a source of single time-bin entangled photon pairs utilizing the biexciton-exciton cascade in a III/V self-assembled quantum dot. We analyzed the generated photon pairs by an inherently phase-stable interferometry technique, facilitating uninterrupted long integration times. We measured visibilities of 78.94(2)~\%, 41.40(3)~\%, and 37.79(3)~\% in three orthogonal bases  and confirmed the entanglement by performing a quantum state tomography of the emitted photons, which yielded a fidelity of 0.69(3) and a concurrence of 0.41(6).}

A source of entangled photon pairs is essential for quantum communication \cite{Duan} and linear optics quantum computing \cite{Knill}. Quantum information protocols such as quantum teleportation \cite{Bouwmeester, Riedmatten1} and entanglement swapping \cite{Pan, Riedmatten2} use entangled photons to enable long distance distribution of entanglement through quantum-repeaters \cite{Briegel}. Optical fibers are the medium of choice for distribution, with the existing extensive global telecommunication fiber network. Polarization entangled photons suffer from decoherence in optical fibers due to polarization mode dispersion \cite{Brodsky, Antonelli}. This effect results in the wavelength and time dependent splitting of the principle states of polarization with a differential group delay. Thus the arrival time of the photons carries information about their polarization state causing decoherence. Alternatively, time-bin entangled photons are immune to these decoherence mechanisms and are more robust in optical fibers \cite{Dynes}. At present, spontaneous parametric down conversion is widely used as a source of time-bin entangled photons \cite{Brendel}. Nevertheless, photons produced by this method show thermal statistics \cite{WallsMilburn} with a finite probability to produce multiple pairs per pump pulse, which limits their performance in quantum information processing \cite{Marcikic}. Thus, there is a requirement for a single photon source \cite{Ana} of time-bin entangled photons. In this work we show that the photon cascade in a single photon pair source such as a single semiconductor quantum dot can fulfil this requirement. 

So far, experimental efforts have been focused on utilizing the biexciton-exciton cascade of a semiconductor quantum dot as a source of polarization entangled photons \cite{Stevenson, Akopian, Dousse}. A major hurdle in the realization of these sources comes from the asymmetry of the self-assembled quantum dots that results in non-degenerate exciton polarization states, thereby revealing the polarization state of the emitted photons. This hurdle has been overcome by some groups with great effort \cite{Stevenson, Akopian, Dousse, Ghali, Kuroda, Muller, Trotta}, but these technologies are not generally available.

\begin{figure}[!ht]
\includegraphics[width=0.5\linewidth]{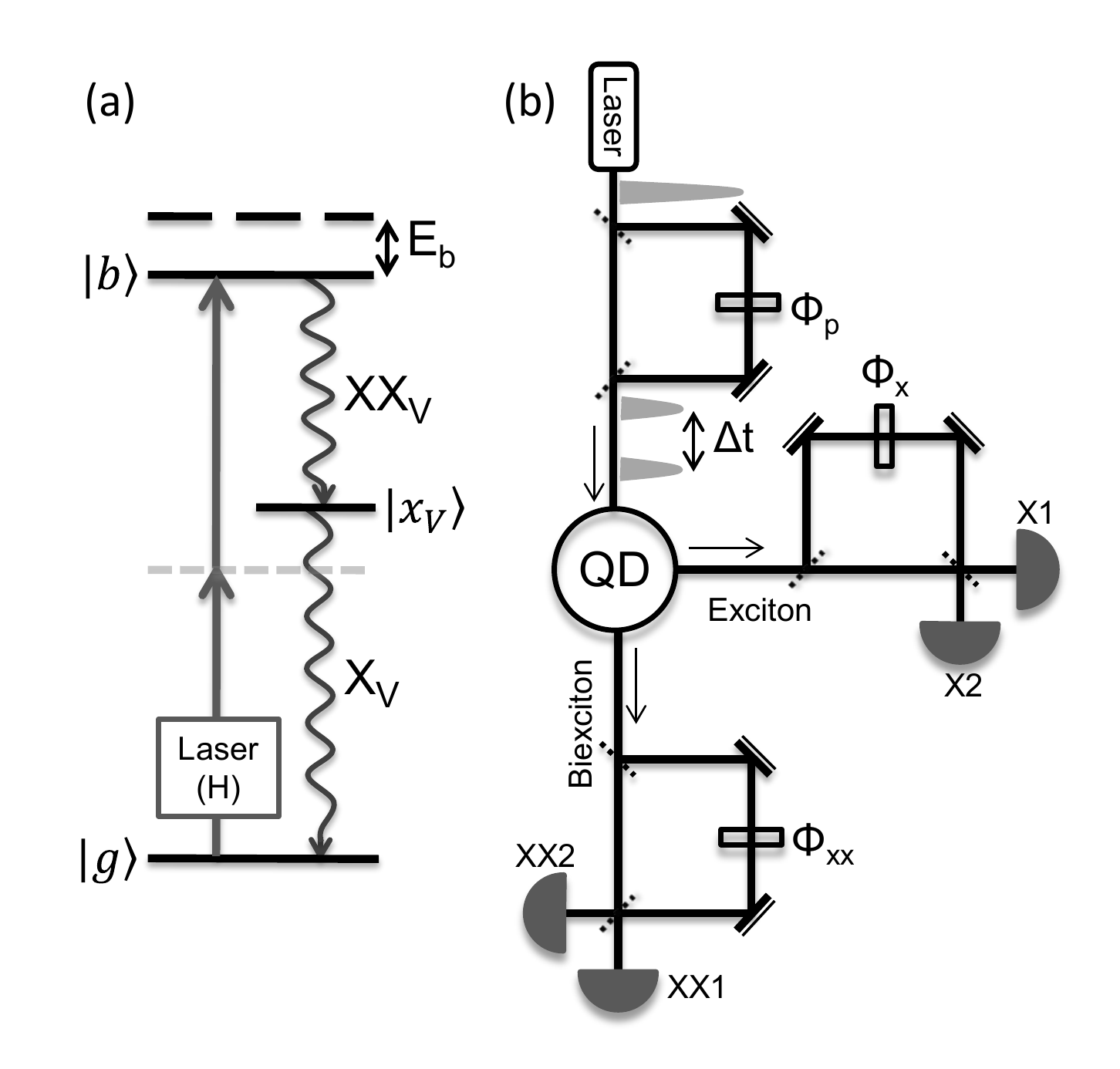}
\caption{\textbf{Schematic of time-bin entanglement generation and analysis.} (a) Energy level scheme for resonant two-photon excitation of a biexciton ($|b\rangle$) from the ground state ($|g\rangle$). H-polarized pump laser populates the biexciton state resonantly through a virtual level (dashed gray line). Following the excitation, a V-polarized biexciton ($XX_V$) exciton ($X_V$) photon cascade is generated through the intermediate exciton state ($|x_{V}\rangle$). Exciton and biexciton photons are detuned from the excitation laser due to the biexciton binding energy $E_b$. (b) Scheme to generate and analyze time-bin entangled photons. Outputs of analyzing interferometers, X1, X2, XX1, and XX2 are fiber coupled to avalanche photon diodes (APD). $\phi_P, \phi_X,$ and $\phi_{XX}$ are the phase in the pump and analyzing interferometers. }
\label{scheme}
\end{figure} 

Our entanglement scheme is insensitive to the polarization non-degeneracy, nevertheless it uses only one exciton polarization cascade so that the emitted photons are in a well defined polarization mode. This method combines the strength of a quantum dot as a single photon source and the robustness of time-bin entanglement. Schemes to generate single pairs of time-bin entangled photons using the biexciton-cascade in a quantum dot have been proposed \cite{Simon, Pathak}. While these schemes require the quantum dot to be initially prepared in a long lived metastable state, we have implemented time-bin entanglement through resonant excitation of the biexciton (Fig.~\ref{scheme}(a)) from the ground state. This method can produce maximally entangled states but it does not allow deterministic creation of entangled photon pairs. The schematic of time-bin entanglement generation from a quantum dot is shown in Fig.~\ref{scheme}(b).

\begin{figure}[!ht]
\includegraphics[trim = 10mm 16mm 0mm 10mm, clip=true,width=0.5\linewidth]{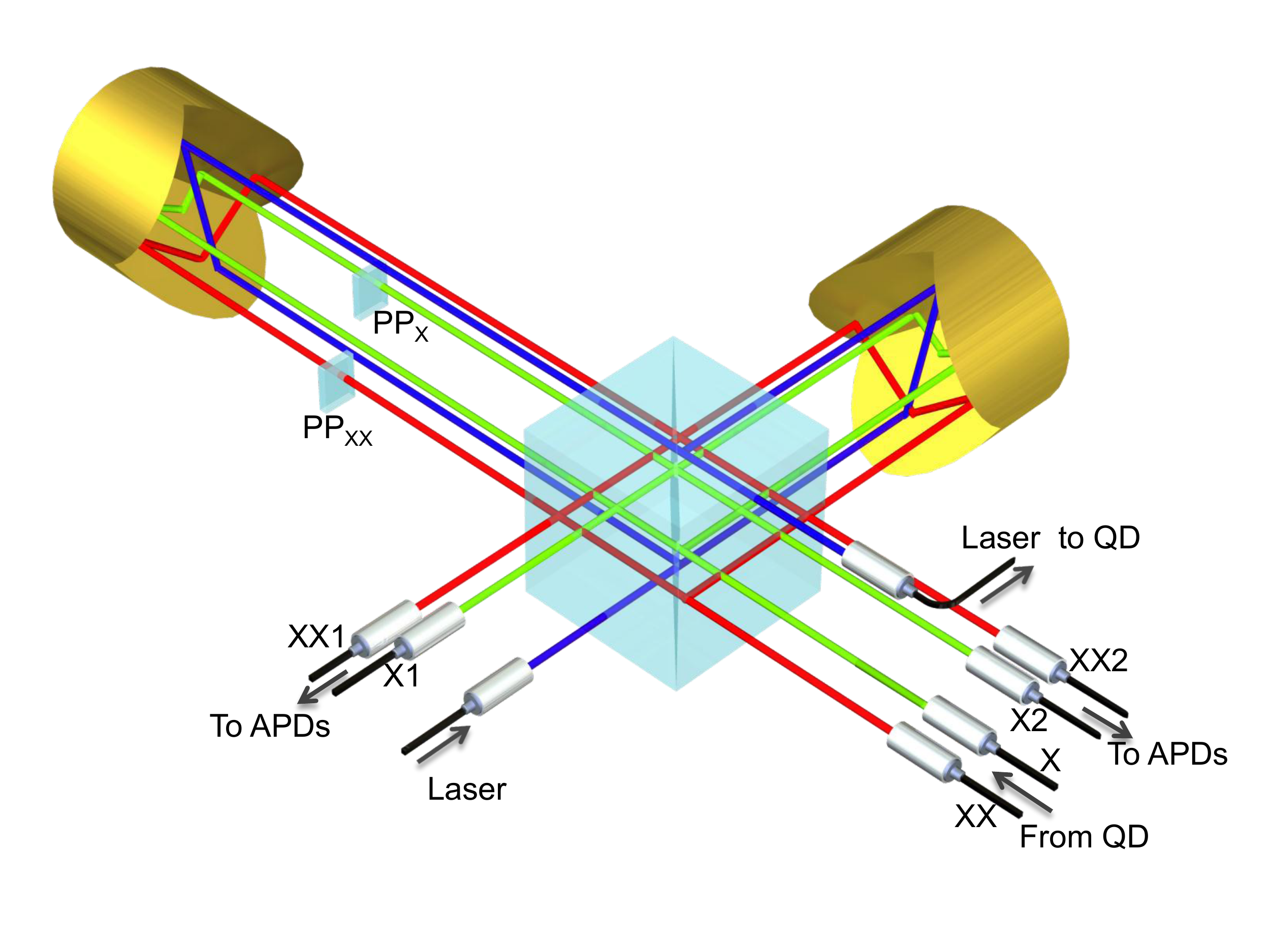}
\caption{\textbf{One bulk interferometer hosts both the pump and analysing interferometers.} The interferometer is built with a 50:50 beam splitter cube and two retroreflectors. The pump laser and the single photon pairs are in separate spatial modes. In and out-coupling of photons in the interferometer is made via single mode fibers. The phase of the biexciton and exciton analyzing interferometers are controlled with phase plates, $PP_{XX}$ and $PP_{X}$.}
\label{setup}
\end{figure}

To excite the quantum dot we use a pump interferometer which transforms the incoming laser pulse into a coherent superposition of two pulses that form well defined time bins ``early'' and ``late''. The Rabi frequency of the excitation laser field at the output of the pump interferometer is given by
\begin{equation}
\Omega_{P}(t)=e^{-i\omega_0 t}[\theta_1f(t)+e^{i\omega_0\Delta t}\theta_2f(t-\Delta\textnormal{t})].
\end{equation}
Here $\theta_1$ and $\theta_2$ are the pulse areas of the two pump pulses with $f(t)$ and $\omega_0$ as the pulse envelope and carrier frequency. The delay in the pump interferometer is set such that the time difference ($\Delta$t = 3.2 ns) between the early and late bins is longer than the width of the laser pulse (4 ps) and coherence time of the biexciton (211 ps) and exciton (178 ps) photons. These laser pulse pairs excite the quantum dot biexciton state through the resonant two-photon excitation process shown in Fig.~\ref{scheme}(a). Details of the excitation process and experimental set-up can be found in \cite{Jayakumar}. The relative phase between the pump pulses creating two-photon excitations is given by $\phi_P=E_{XX}\Delta t/\hbar$ where $E_{XX}$ is the biexciton photon energy. Resonant excitation is crucial as it coherently transfers the phase of the pump pulses and thus the coherence created by the pump interferometer to the biexciton state. The coherence of the excitation process was proved in our previous work \cite{Jayakumar}, through coherent manipulation experiments demonstrating Ramsey interference. In contrast other, non-resonant excitation techniques involve phonon transitions, which leak the creation time of the biexciton to the environment \cite{Roszak}, thereby degrading the coherence. At sufficiently low excitation power the biexciton state is created either by the early or the late pulse, followed by the emission of a biexciton-exciton photon cascade. The emitted photons are in the time-bin entangled state 
\begin{equation}
|\Phi\rangle=\frac{1}{\sqrt2}(|early\rangle_{xx}|early\rangle_{x}+e^{i\phi_{p}}|late\rangle_{xx}|late\rangle_{x}),
\end{equation}
where the photon pairs are in a coherent superposition of being emitted in the early or late time bin. We analyzed the entanglement of the emitted photons in a Franson-type two-photon interference experiment \cite{Franson} as shown in Fig.~\ref{scheme}(b). The exciton and biexciton photons were fed into two separate analyzing interferometers that had delays equal to the pump interferometer. Finally we recorded the coincidences between the outputs of the analyzing exciton and biexciton interferometers. Photon pairs produced by the early pulse and taking the long path in the analyzing interferometers are indistinguishable from photon pairs produced by the late pulse and taking the short path. Thus, the probability amplitudes of the two possible indistinguishable events interfere. This interference results in a sinusoidal change of the coincidence rates while the phase of one of the interferometers is varied. The relationship between coincidence counts and interferometer phase can be approximated \cite{Franson} by 
\begin{equation}
R \sim  1 - ij V \cos(\phi_{XX}+\phi_{X}-\phi_{P}),
\label{visibility}
\end{equation}

with $i$, $j$ taken as $+1(-1)$ for the XX and X outputs $1(2)$, respectively. The visibility ($V$) of the interference is connected to the quality of the time-bin entanglement and it is ideally equal to unity. In order to achieve maximum visibility the phase ($\phi_{XX}+\phi_{X}- \phi_{P}$) has to be stable during the experiment. This can be accomplished by active stabilization of all three interferometers with respect to a phase-stable laser. In our experiment, the phase stability is achieved by realizing the pump and analyzing interferometers in three different spatial modes of a single bulk interferometer as shown in Fig.~\ref{setup}. We take advantage of the phase relation, such that the relative phase always remains the same independent of the drift in the interferometer. The phase of both the analyzing interferometers can be set independently with the phase plates $\textnormal{PP}_{X}$ and $\textnormal{PP}_{XX}$. 

\begin{figure}[!ht]
\includegraphics[width=1\linewidth]{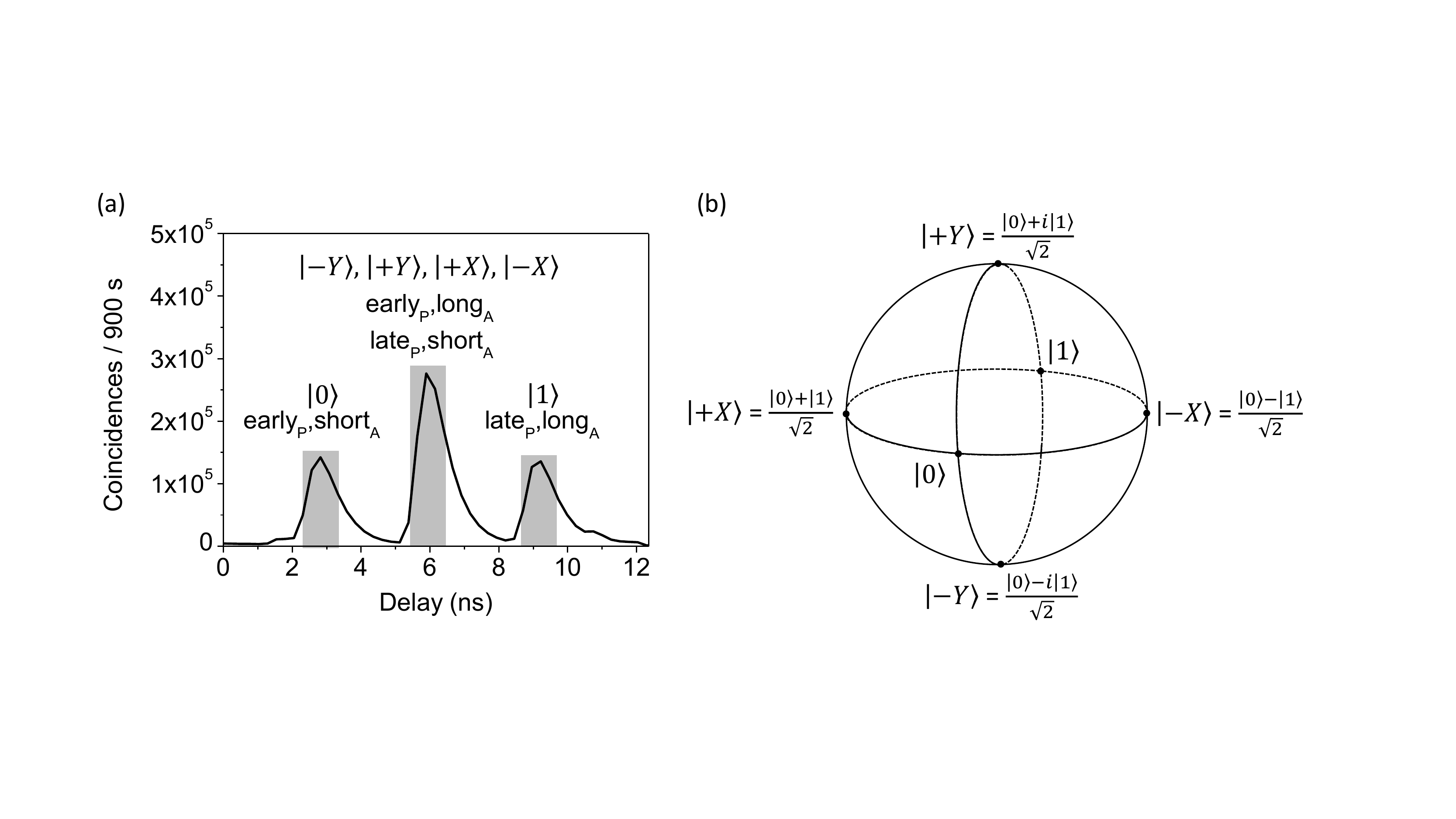}
\caption{\textbf{Time-bin qubit and Bloch sphere representation.} (a) Arrival time of exciton photon recorded with respected to the pump pulse. Gray shaded regions are the 1.28~ns windows taken for extracting the triple coincidence (laser-biexciton-exciton). Side peaks are the projections to the time basis $|0\rangle$and $|1\rangle$. Coincidences in the middle are projections on the energy basis  $|X\rangle$ and $|Y\rangle$ for phase settings $\phi_X=0, \pi/2$. (b) Qubit states represented on a Bloch sphere.  }
\label{threepeaks}
\end{figure}

\begin{figure}[!ht]
\includegraphics[trim = 0mm 0mm 0mm 0mm, clip=true,width=0.5\linewidth]{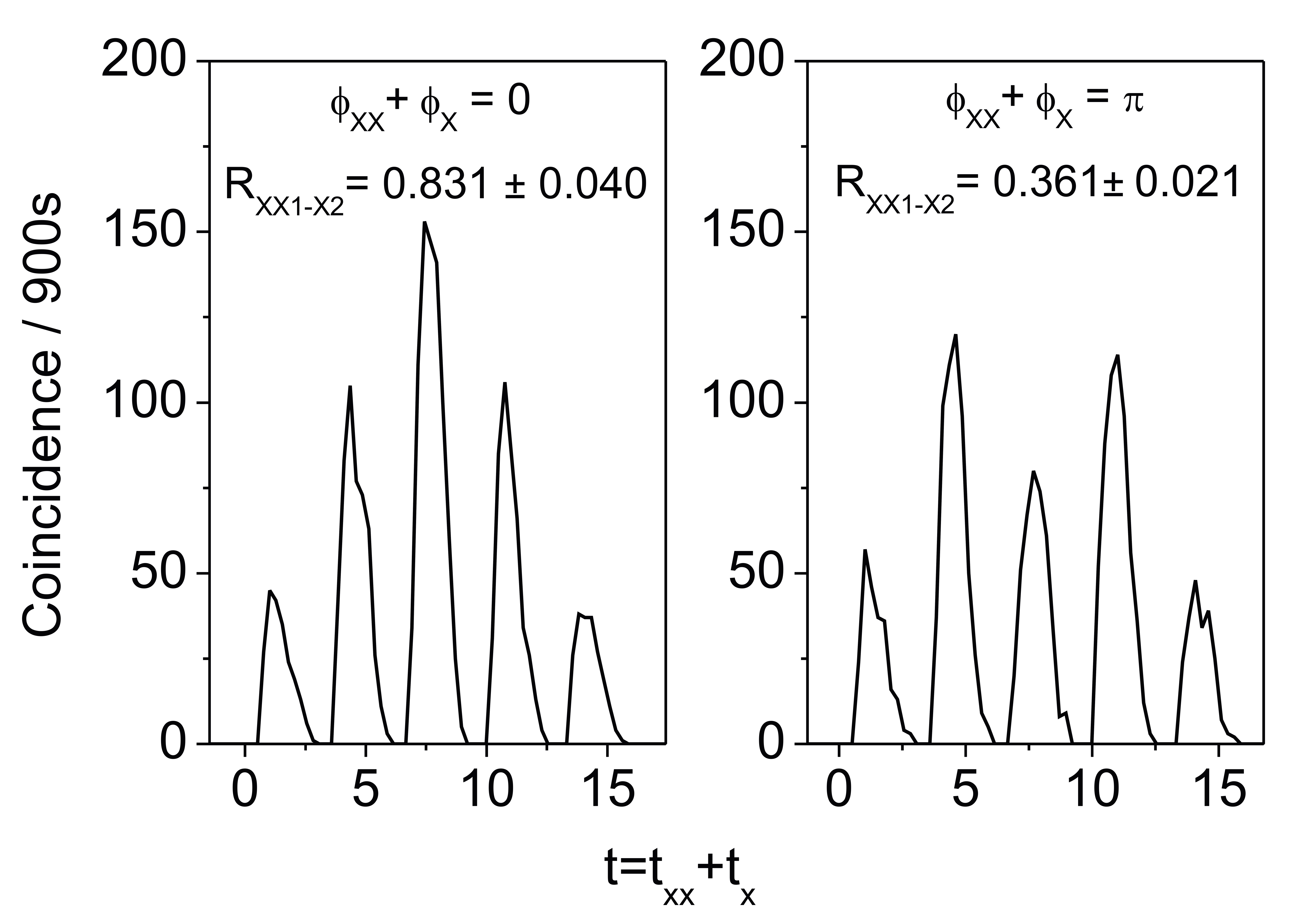}
\caption{\textbf{Triple coincidence.} Coincidences between the laser pulse, biexciton (XX1) and exciton (X2) outputs of the analyzing interferometers. Triple coincidences recorded for phase setting $\phi_{XX}+\phi_{X}$ = 0 and $\phi_{XX}+\phi_{X}$ = $\pi$. The arrival time of the biexciton and exciton photon at the output of the analyzing interferometers with respect to the laser pulse is represented by $t_{XX}$ and $t_{X}$. Normalized coincidences for each phase setting is shown.}
\label{fivepeaks}
\end{figure}

In order to extract coincidences between the photons at the interferometer outputs, the arrival times of the photons at the output of the analyzing interferometers were recorded with respect to the pump pulse.  For example, Fig.~\ref{threepeaks}(a) shows the recorded events for output X1. The photons created by the early (late) pulse and traveling the short (long) path form the first (third) peak. The second peak is formed by the photons created in the early (late) pulse and taking the long (short) path. Photons arriving within a window of 1.28~ns in each of the peaks, shaded with gray in Fig.~\ref{threepeaks}(a) were post-selected. From the post-selected events, coincident events were extracted between the exciton and biexciton outputs. The extracted triple coincidence histogram has five peaks as shown in Fig.~\ref{fivepeaks}. The first (last) peak consists of the pairs created by the early (late) pulse and passing through the short (long) paths of the analysing interferometers.  The second (fourth) peak consists of the pairs created by the early (late) pulse and one of the emitted photons taking the long path. The third peak contains the indistinguishable events that interfere and exhibit the entanglement. The triple coincidence between the laser, biexciton and exciton photon is required to separate the coincidences in the third peak (interfering) from the coincidences of the first and last peaks.  

We quantified the time-bin entanglement of the emitted photon pairs by measuring their quantum state through quantum state tomography \cite{Takesue}. With this method we reconstructed the density matrix of the generated time-bin entangled photons. We recorded coincidence events for different projections of the individual qubits (exciton and biexciton photons) to the states $|0\rangle$, $|1\rangle$, $|+X\rangle$ and $|+Y\rangle$  (represented on the Bloch sphere shown in Fig.~\ref{threepeaks}(b)). 

Projections onto the states  $|X\rangle$ and $|Y\rangle$ (energy basis) are achieved by setting the phase in the analyzing interferometer to 0 and $\frac{\pi}{2}$, respectively.

It takes four measurements (see Methods) to obtain all sixteen projections between the exciton and biexciton qubits required for the state reconstruction  \cite{Takesue}. 

\begin{figure}[!ht]
\includegraphics[trim = 30mm 10mm 30mm 0mm, clip=true,width=0.5\linewidth]{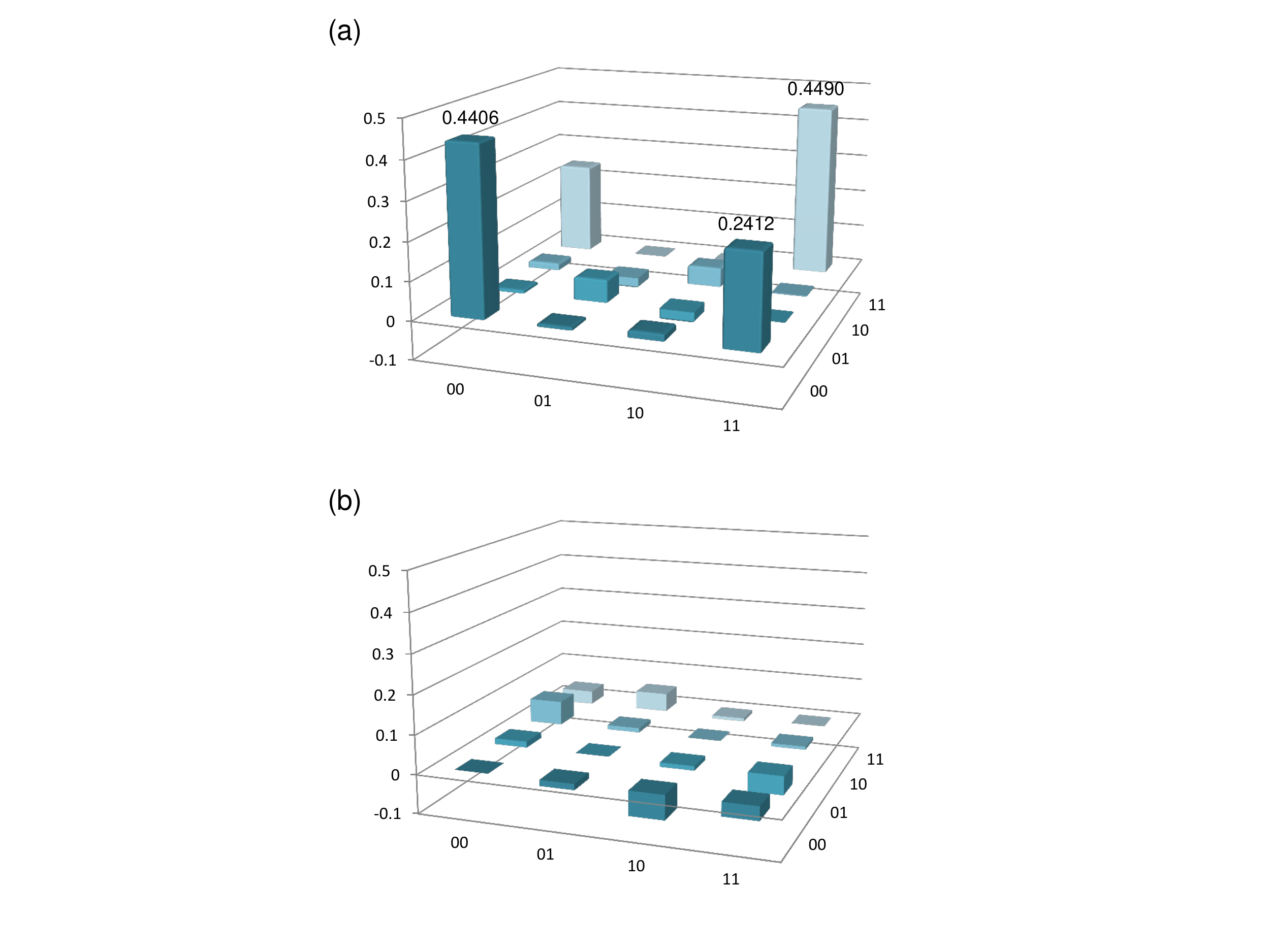}
\caption{\textbf{Density matrix for XX1-X2 outputs plotted as a bar chart.} (a) Real part. (b) Imaginary part. The reduced value of 0.44 for the 00 (early-early) and 11 (late-late) populations is a consequence of the double excitations that appear as 01 (early-late) and 10 (late-early) populations. The reduced value of the coherence, 0.25, comes from the lack of transform-limited coherence in the quantum dot.}
\label{density}
\end{figure}

The analyzing interferometers phase settings for these four measurements are $(\phi_{XX},\phi_{X})$ = $ (0,0)$, $(\frac{\pi}{2},0)$, $(0,\frac{\pi}{2})$ and $(\frac{\pi}{2},\frac{\pi}{2})$. 
The real and imaginary parts of the reconstructed density matrix are shown in Fig.~\ref{density}, from which we obtain a fidelity of 0.69(3) with respect to the $|\Phi^+\rangle$ state, a tangle of 0.17(5) and a concurrence of 0.41(6).

The visibility of the two-photon interference was obtained from the coincidence events recorded with the phase settings, $(\phi_{XX},\phi_{X})$ = $ (0,0)$ and $(\frac{\pi}{2},\frac{\pi}{2})$. Fig.~\ref{fivepeaks} shows the triple coincidences between the biexciton and exciton outputs of the analyzing interferometers (XX1 and X2  in Fig.~\ref{setup}). 
%Despite recording the coincidences for a fixed time duration for the two phase settings, the coincidence counts of the interfering peak can not be directly compared between measurements taken at different times, because the rates vary as a result of the drifting of the cryostat. These variations were compensated by normalizing the coincidence counts of the interfering peak with the coincidences of peaks two and four $(R=2A_3/(A_2+A_4)$, where A is the total count of the respective peak). The normalized coincidence values are shown in Fig.~\ref{fivepeaks}. 
The visibilities in three orthogonal bases $0/1$, $+X/-X$, and $+Y/ -Y$ were measured to be 78.94(2)~\%, 41.40(3)~\%, and 37.79(3)~\%, respectively.
%for XX1-X2 outputs. Triple coincidence and visibility values for other output pairs are shown in the supplementary material. 

The imperfect interference visibility and hence the reduced fidelity of the entangled state can be attributed to the following reasons. Ideally, we expect pairs to be produced in only one of the two pulses. For the excitation power used in this experiment, we estimate the coincidences between the pairs created by the two pulses to be 12.4\% of the coincidences from the same pulse. These coincidences between the two pairs form an incoherent background, hence reducing the visibility. The choice of the pulse power was limited by the photon pair collection and detection efficiency. We also attribute some decoherence of the entangled state to the pure dephasing of excitons and biexcitons in the solid state environment due to phonon interactions and spectral diffusion. These interactions leak information about the creation time of the biexciton \cite{Roszak} and energy fluctuations of the excitons reduce the visibility of two-photon interference. The dephasing in the quantum dot used for the experiment was characterized with lifetime (biexciton: 405 ps , exciton: 771 ps) and coherence length (biexciton: 211 ps , exciton: 178 ps) measurements which is far from the ideal transform limited condition. 

We have generated time-bin entangled photons from a quantum dot using the biexciton-exciton cascade. Entanglement in the generated photons was confirmed by reconstructing the density matrix through quantum state tomography. This opens up opportunities to use quantum dots as a source of robust entangled photon pairs that is not affected by fine structure splitting. The entanglement of our source can be improved by reducing the generation of photon pairs in both the excitation pulses by reducing the excitation laser power. This can be achieved with quantum dots in coupled pillar microcavities \cite{Dousse} capable of high extraction of both exciton and biexciton photons through Purcell enhancement. Purcell enhancement will also improve the entanglement by the emission of photon pairs that are close to the Fourier transform limit. The entangled photon pairs can be created deterministically by identifying a metastable state \cite{Simon, Pathak}. 

\section*{Methods}

A measurement of the time of arrival of a photon at the output of the analyzing interferometer with respect to the excitation laser gives rise to three peaks shown in Fig.~\ref{threepeaks}(a). Here, the first and the last peak carry the information on classical correlations (between $|0\rangle$ and $|1\rangle$) and the middle peak shows quantum correlations. For example $|+Y,0\rangle$ is a projection where the biexciton photon is projected onto the energy basis and the exciton is projected onto the time basis. The joined projections onto the time basis ($|0,0\rangle, |0,1\rangle, |1,0\rangle, |1,1\rangle$) are obtained as coincidence events in respective combination of the time intervals measured between the outputs of the analyzing interferometers for biexciton and exciton. Projections onto the energy basis ($|X\rangle, |Y\rangle$) are obtained from respective specific phase settings by extracting coincidence events in the time of arrival of the middle peak for exciton and biexciton signal. Projections with energy and time basis combinations ($|0,+Y\rangle, |+Y,0\rangle, |+Y,1\rangle, |1,+Y\rangle$, $|0,+X\rangle, |+X,0\rangle, |1,+X\rangle, |+X,1\rangle$) are collected from coincidence events in the interval of arrival of the respective middle and side peaks. In order to obtain the measurement errors we performed a 100 run Monte Carlo simulation of the data with a Poissonian noise model applied to the measured values.

\section*{Acknowledgements}
This work was funded by the European Research Council (project EnSeNa) and the Canadian Institute for Advanced Research through its Quantum Information Processing program. G.S.S.
acknowledges partial support through the Physics Frontier Center at the Joint Quantum Institute (PFC@JQI).

\end{document}